# Formation of subject area and the co-authors network by sounding of Google Scholar Citations service

**D.V. Lande, V. B. Andrushchenko**

**Institution for Information Recording of National Academy of Science of Ukraine**
*Ukraine, Kyiv*
dwlande@gmail.com, valentyna.andrushchenko@gmail.com

The suggested methodic is the way of formatting the subject areas models and co-authors networks by sounding the content networks. The paper represents the notion networks which match tags and authors of Google Scholar Citations service. Models depicted in the work were built for the physical optics area, and it can be applied for other domains. The proposed ways of defining connections between science areas and authors depicts the collaborations opportunities and versatility of interdisciplinary.

Today science and technique progress depends heavily on the way scientists can establish the right cooperation process and create the successful collaboration. And activity in every science field assumes not only experiments and their depiction in scholar papers, but the repercussion of scientific results. And first of all - the way the paper is cited, when, whom and in what way. The information scientometric databases allow researchers trace their activity and fix their achievements, and this data is actual not only for the scientific image but for the way of promoting their research all over the world and even for career.

New information resources give new opportunities for describing the subject areas and studying the consistent pattern of the scientific intercommunications.

One of the main instruments of investigating the regularity of scientific cooperation - is the co-authorship network forming by scientometrics services [2].

Co-authorship network permits to:

- obtain scientometric indexes;
- find experts to solve complex problems;
- define the participants for the expertise procedure;
- search out colleagues for the collaboration formation;
- discover interconnections between scholars for further scientometric analysis;
- provide the estimations for the grant effectiveness and the performance of research institution;
- etc.

There are some major services of scientific information, which give an opportunity to get the scientometric data, create users profiles and can also contain bibliographic information, these databases are Web of Science and Scopus. But the access to the resources of these services is paid and it complicates the work with them.

One of the main services with the free access is Google Scholar, which allows creation of profile containing information about all the published materials of the researcher, and also has the powerful search engine.

Authors experienced the approach to form the subject domain model and create the co-authors network (‘text mining’ [3] ‘legal science’ [5], ‘physical optics’) by sounding big information network and creating the notion network, matching with tags of sceintometric service Google Scholar Citations (http://scholar.google.com/citations) [3], [4].

The interface of the Google Scholar Citations gives the authors lists and the complemented tags (notions and concepts), which are corresponded with the preassigned tag.

The work also presents the algorithm of co-authors network formation - the model of researchers cooperation by sounding mentioned scientometric network.

The proposed sequence of operations according to the algorithm [1]::

1. The short list of the base tags, defined in an expert way.
2. One tag is chosen from the list.
3. The performed search represents the web-page corresponding with the chosen tag.
4. The neighboring tags, comprised in the page, are added to the forming network.
5. From the neighboring tags are choses the ones, pages of which are planned to be proceed for the further analysis. This tag is the one with the highest rate, which responds the theme of the chosen subject area, and the transition to which hasn’t been executed.
6. If such a tag been chosen the jump to the step 3 is to be provided.
7. If there is no such a tag, and the list of base tags is not fulfilled, then the segue to the next base tag from the initial tag list is provided (Step2). Otherwise the network is considered to be built.

The described algorithm was complemented by the rule, which reads that the Steps 3-6 were provided so many times as expert set. In example with the ‘physical optics’ tag it makes up 5.

Picture 1 shows the example of the subject area notion network, built according to the algorithm on base tags. Picture 2 - the enlarged part of the network.

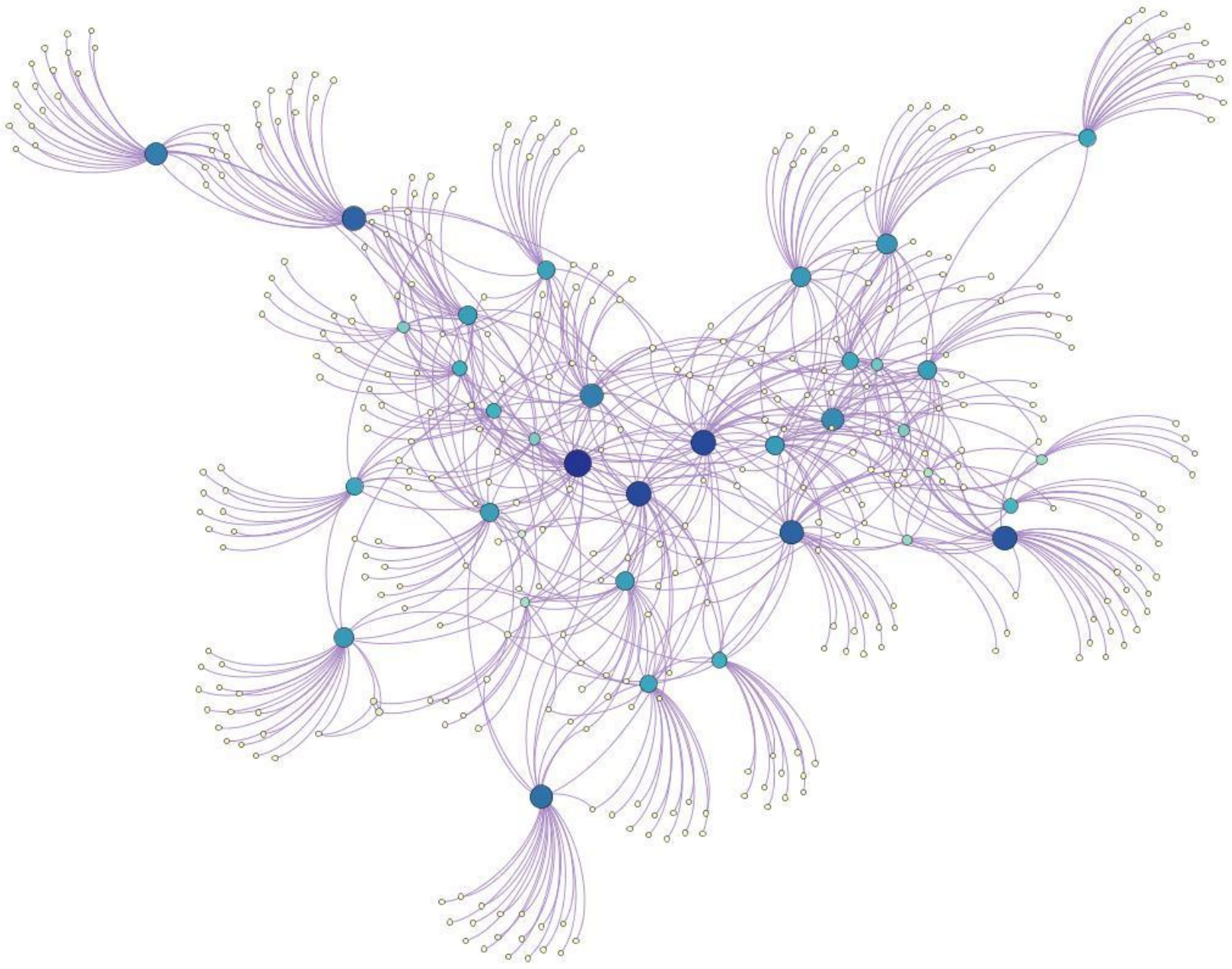

Picture. 1. The structure of the network according to the base tag

By scanning the Google Scholar Citations there was obtained the network with such parameters:

- the nods number - 401;
- the edges number - 670;
- the average node rate - 1,67;
- the diameter - 6;
- the average shortest way - 2, 48.

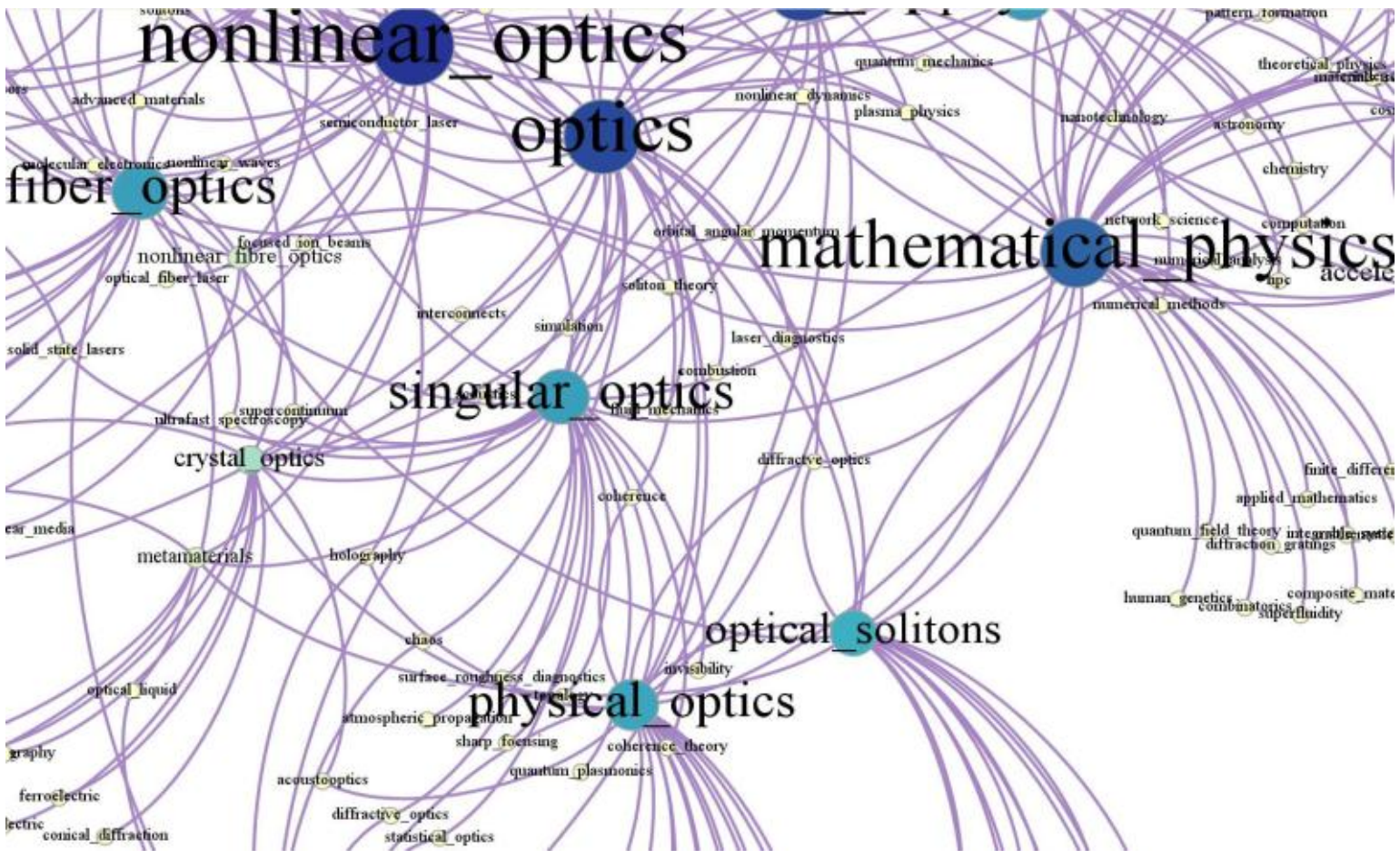

Picture. 2. The part of the notion network

The given principle was adopted to the formation of co-authorship network [3]:

1. The base tag is expertly defined (as an example - ‘physical optics’) and there are several further steps:
2. The page, corresponded to this page opens.
3. The most cited author, represented on the page is chosen.
4. All the authors from the chosen author’s page are added to the framing network. The edges-connections to these nodes (co-authors) form the initial nod (author).
5. From the list of nods the one is chosen, transition on which page is planned for the further analysis. This is the most weighty nod responded the theme of chosen subject domain (its tags contain the parts of words, singled out by experts, ‘physic’ and ‘optics’ in particular) and it is not the one from the nods, which pages were observed.
6. If there is no such an author the network is considered to be completed.

In appliance with the algorithm the co-authors network was built taking into the consideration the restrictions for the number of scanning nodes. The route of the algorithm, the list of researchers and their tags are depicted on the picture 3.

As the result of the scientometric network sounding there was obtained the co-authors network with the next parameters:

- the nods number - 207;
- the edges number - 384;
- the average node rate - 2,34;
- the diameter - 7;
- the average shortest way - 4,23.

**Tiberiu Tudor** physical_optics polarization coherence lasers quantum_optics
**Sabino Chavez-Cerda** optics mathematical_physics physical_optics diffractve_optics optical_solitons
**David Snchez-de-la-Llave** optics physical_optics fourier_optics_and_signal_processing holography
**Miguel A. Bandres** physics optics photonics
**Johannes Courtial** physics optics ray_optics holography
**Mark R Dennis** mathematical_physics optics singular_optics topology
**Franco Nori** condensed_matter_physics quantum_optics quantum_information physics superconductivity
**Gran Johansson** quantum_physics quantum_computing microwave_quantum_optics the_dynamical_casimir_effect mesoscopic_superconductivity
**Abraham G. Kofman** quantum_physics quantum_information quantum_optics laser_physics solid_state_qubits
**Skab Ihor** physical_optics singular_optics crystal_optics piezo_and_electrooptics acoustooptics
**Eduard Carcol''** physical_optics seismology computers
**Neill Lambert** physics quantum_optics quantum_computing nano_mechanics quantum_mechanics
**Arend G. Dijkstra** theoretical_chemical_physics nonlinear_optics open_quantum_systems
**B. M. Rodrguez-Lara** quantum_optics optical_physics
**Suren A. Chilingaryan** quantum_optics_and_quantum_information quantum_physics quantum_mechanics
**Myun-Sik Kim** metrology interferometry physical_optics phase_anomaly microlens
**G. Rodriguez Zurita** physical_optics interferometry fourier_optics
**Vlokh Rostyslav** physical_optics
**Karol Bartkiewicz** quantum_physics quantum_optics quantum_information
**Anirban Pathak** physics quantum_information quantum_optics
**Swapan Mandal** quantum_optics laser_spectroscopy quantum_information_theory mathematical_physics
**Ioannis Besieris** stochastic_linear_and_nonlinear_wave_propagation phase_space_techniques wave_localization

Picture 3. The route of forming the co-authors network

Using the Gephi application the visualization of the network became available (Picture 4). Application of the clustering analysis makes the detection of the closest researchers-co-authors groups, scientific schools, and experts groups approachable.

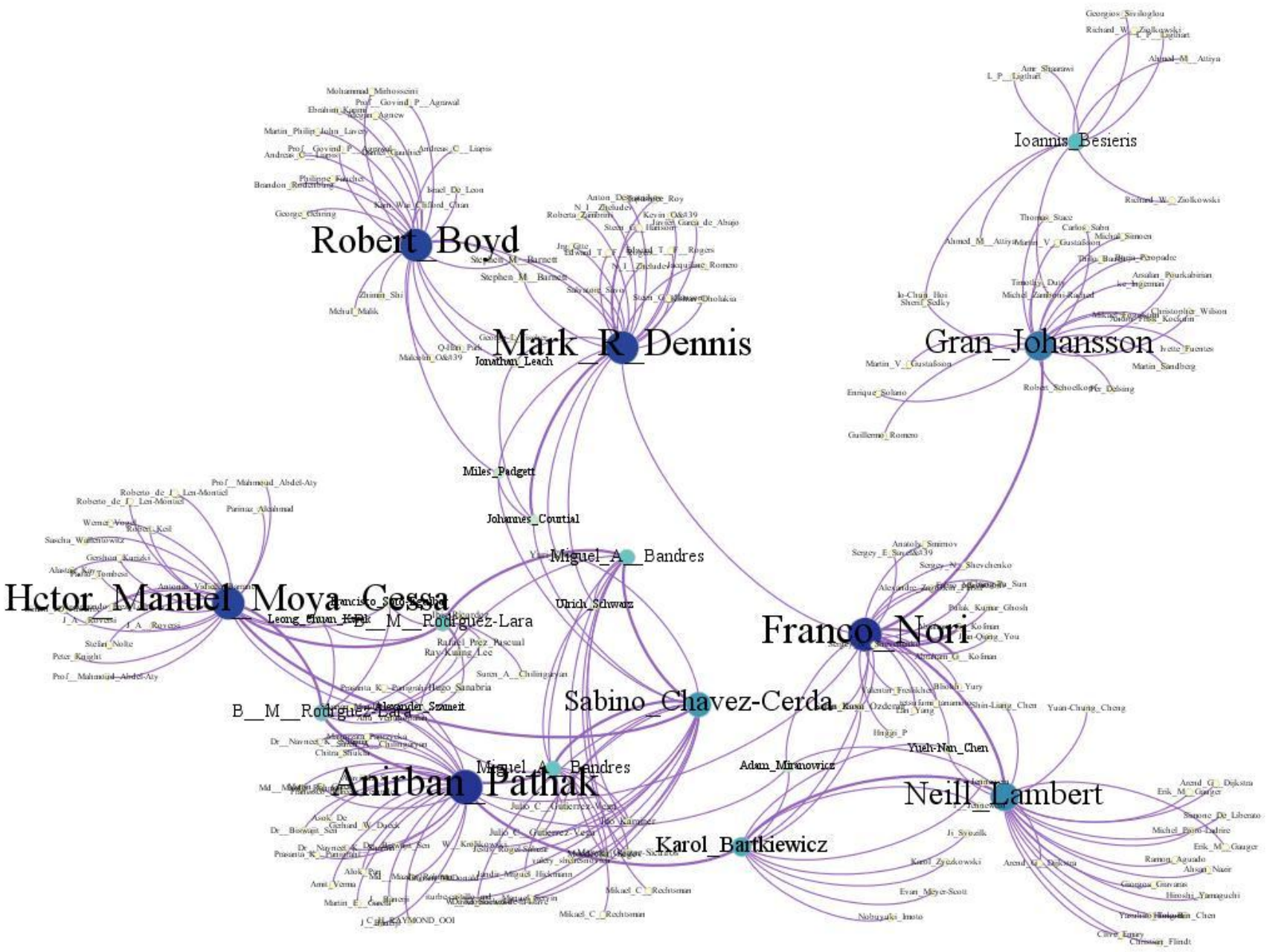

Picture 4. The structure of the co-authors network.

If we leave only structurally weighty nodes and edges, using Gephi, the clustering of the initial network can be reached, and also the most connected subgroups of researchers (Picture 5).

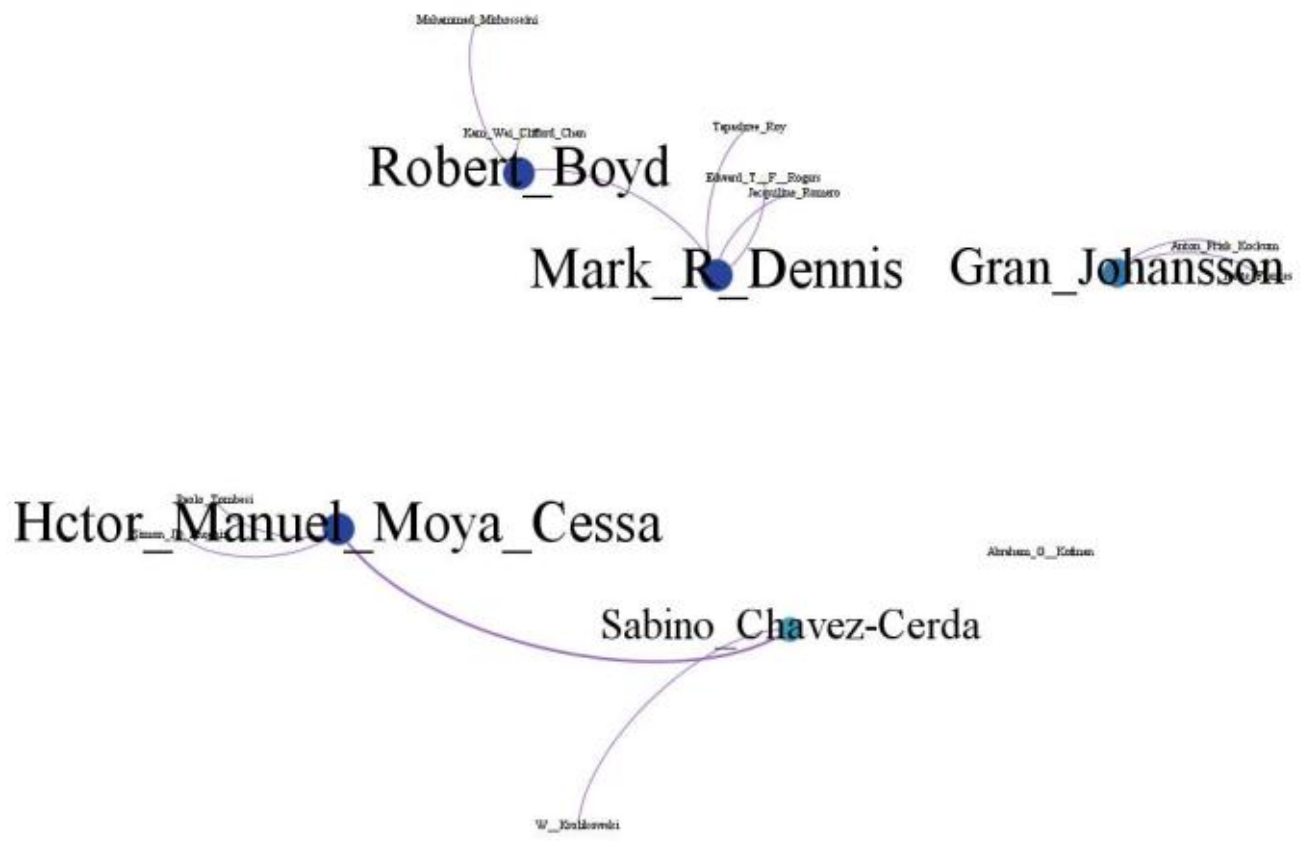

Picture 5. The biggest clusters of the represented network.

Thus the suggested attempt to form the model of subject domain and the co-authors network in frames of the defined subject area, the measured elements of which are knowledge markers (tags), preassigned by researchers-participants of Google Scholar Citations project. It's necessary to notice the fundamental difference of the represented model of automatic subject domain model forming from the existed ones, based on the text corpuses analysis or the direct participation of experts in process of electing nodes and edges. This way the expert-user inputs only the grains of the knowledge as tags and small dictionaries (up to 10 words). Hereinafter program uses information provided by the authors of publications and tags noted as the main ones.

The work is the actual application for the information scientometric databases. It clearly widens the existed facilities and represents the great amount of analytical information vital not only for researchers but for the research institutions as a way of monitoring the dynamics of scholars activity and cooperation, and also can appear the recommended instrument in founding science policy of the country.